\documentclass[%
 reprint,
 amsmath,amssymb,
aps,
prstper,
]{revtex4-2}

\usepackage{graphicx}
\usepackage{dcolumn}
\usepackage{bm}
\usepackage{braket}

\begin{document}

\preprint{APS/123-QED}

\title{Light with Even Fock states from Interference of Kerr-squeezed Light}

\author{Ziv Abelson}
 \email{zivabelson@hotmail.com}
\author{Shimshon Bar-Ad}%
 \email{shimshon@tauex.tau.ac.il}
\affiliation{%
 Sackler School of Physics and Astronomy, Faculty of Exact Sciences, Tel-Aviv University, Tel Aviv 6997801, Israel}

\date{\today}

\begin{abstract}
We demonstrate the generation of non-classical light by destructive interference of identical Kerr squeezed states. Perfect pair-wise cancellation of amplitudes that contribute to odd Fock states results in
light with only even Fock states, independent of the strength of the nonlinearity. The observability of this
effect is only limited by the quality of the interferometer. In the low nonlinearity limit, the even-only state
resembles a squeezed vacuum state, yet the even-odd oscillations persist when the nonlinearity is strong.
The effect is also robust against deviations from the optimum relative phase of the input Kerr-squeezed
states.
\end{abstract}

\maketitle


Non-classical states of light are of interest from a fundamental point of view, and also have technological
applications, especially in the context of precision quantum measurements, which take advantage of the
reduced phase or amplitude uncertainty due to squeezing, and the sub-Poisson photon statistics of amplitude
squeezed states. Such states are typically generated using nonlinear parametric down conversion [\onlinecite{gerry2023introductory,PhysRevLett.59.2566.Slusher,Breitenbach_Homodyne,Wu:87,breitenbach1997measurement}]. Other nonlinear processes, such as four-wave mixing [\onlinecite{PhysRevLett.55.2409}] and second harmonic generation [\onlinecite{PhysRevLett.72.3807}] have also been implemented. Self-phase modulation by a Kerr nonlinearity, which is easy to implement, converts amplitude fluctuations to concomitant phase fluctuations, thereby inducing quadrature squeezing, but the photon number statistics remain Poissonian [\onlinecite{loudon2000quantum}]. Yet, by mixing such a state with coherent light, sub-Poissonian (or super-Poissonian) photon statistics can be obtained. This was shown theoretically [\onlinecite{PhysRevA.34.3974, PhysRevA.41.2906,RITZE1979126,PhysRevA.44.7647}] and demonstrated experimentally [\onlinecite{PhysRevLett.81.2446,Krylov:98,andrianov2023fiber,Kalinin}]. Similar arrangements were also used to generate squeezed optical solitons [\onlinecite{PhysRevLett.66.153}] and a continuous-variable Einstein-Podolsky-Rosen (EPR) entanglement [\onlinecite{PhysRevLett.86.4267}]. Independently, mixing of quantum and classical light allowed the generation of NOON multiphoton entangled states [\onlinecite{doi:10.1126/science.1188172}], which are also useful for precision measurements.

Here we discuss a new mechanism for obtaining non-classical light, with photon number oscillations, involving straightforward Kerr squeezing. We show how destructive interference of identical Kerr-squeezed states leads to light with exclusively even Fock states. This effect is a result of perfect pair-wise cancellation of amplitudes that contribute to odd states, implying that its observability with an arbitrarily low nonlinearity is only limited by the quality of the interferometer. In the low nonlinearity limit the new state resembles a squeezed vacuum state, yet the even-odd oscillations persist when the nonlinearity is strong, in which case they are robust against deviations from the optimum relative phase of the input Kerr squeezed states. To the best of our knowledge, similar even-odd oscillations, namely squeezed vacuum states, have been demonstrated before only by using parametric processes [\onlinecite{breitenbach1997measurement}], while Kerr-Kerr interference was proposed only as a conditional scheme for preparing coherent state superpositions [\onlinecite{adam2011double}].

\begin{figure}[b]
\includegraphics[width=8.8cm]{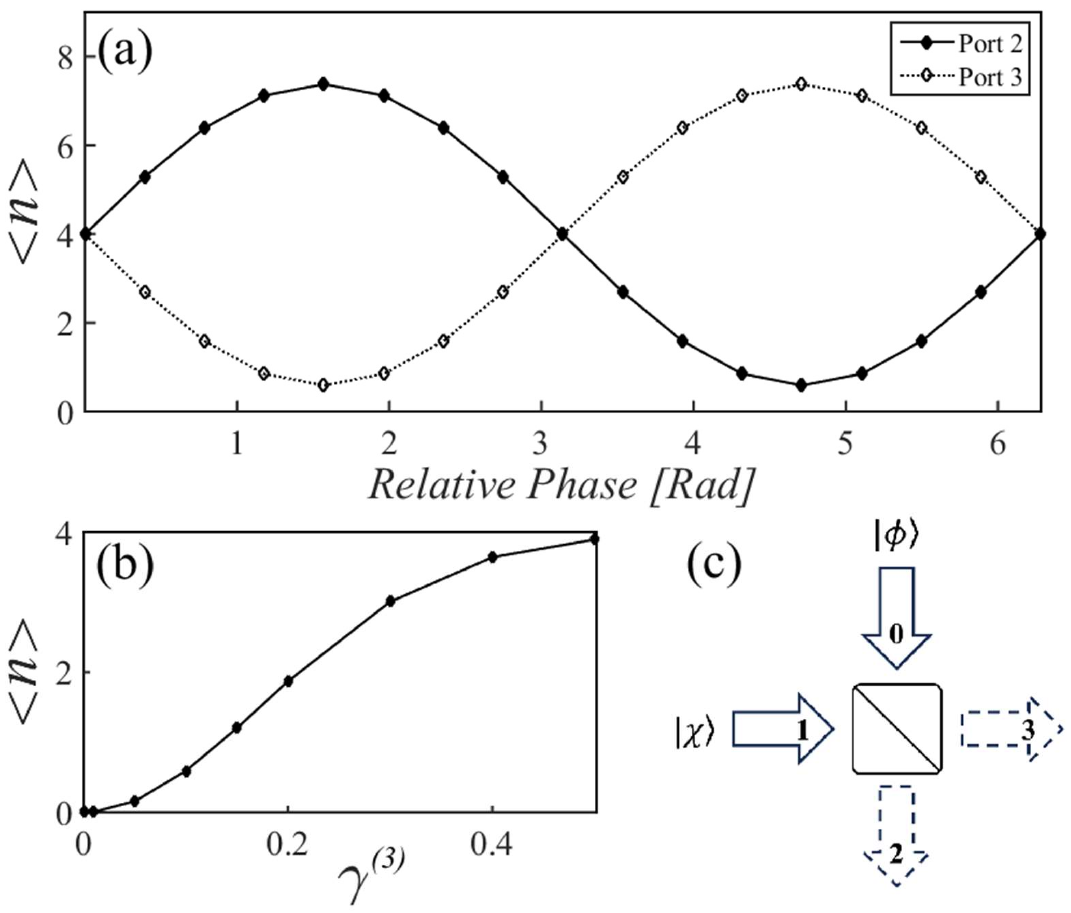}
\caption{\label{fig:n_average} interference of two Kerr states of equal amplitudes. (a) shows numerical evaluations of Eqs. \ref{eq:Eq_2} and \ref{eq:Eq_3} for
a 50:50 beam splitter, \begin{math} \beta=2 \end{math}, \begin{math} \alpha=2e^{-i\theta} \end{math} and \begin{math} \gamma^{(3)}=0.1 \end{math}, as a function of the relative phase \begin{math} \theta\end{math}. (b) shows the average
photon number exiting port 3 as a function of \begin{math} \gamma^{(3)} \end{math}, for \begin{math} \theta=\pi/2 \end{math}. (c) shows a schematic of the interference setup.
}
\end{figure} 

We demonstrate the odd-even oscillations in the framework of a fully quantum, single-mode model, which describes the interference in a general beam splitter, using the density matrix formalism. The input state in the calculation is \begin{math} \ket{\phi}_0\ket{\chi}_1 \end{math} where \begin{math} \phi \end{math} and \begin{math} \chi \end{math} denote coherent and/or Kerr squeezed states (namely coherent states that have undergone self-phase modulation in a Kerr medium), and 0,1 denote the input ports of the beam splitter (see Figure \ref{fig:n_average}(c)). The statistical properties (photon number and phase distributions) of the states at the two output ports are calculated as a function of the relative phase of the two input states.
\begin{figure*}[t]
\includegraphics[width=18.3cm]{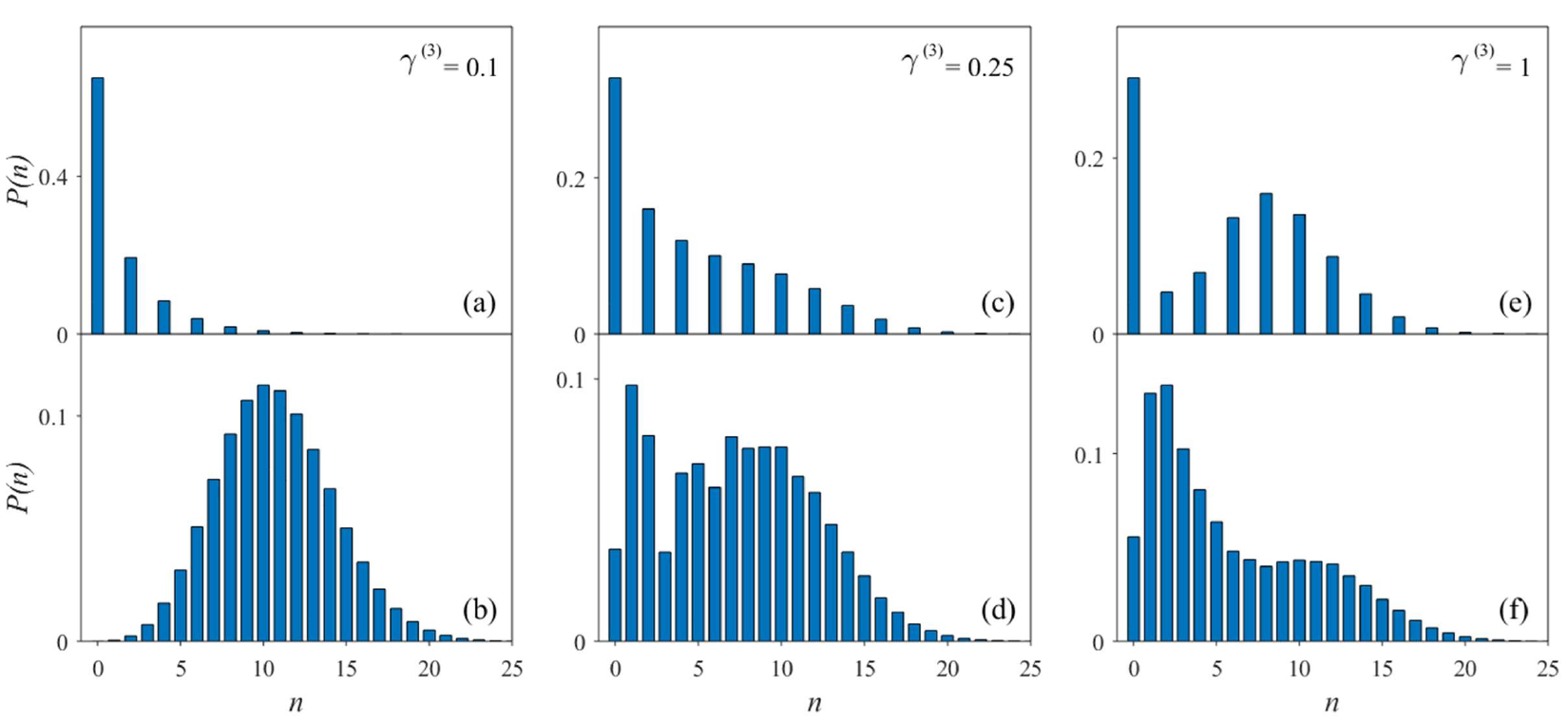}
\caption{\label{fig:with_increasing_gamma} the calculated photon number distributions at output ports 3 (top) and 2 (bottom) of a symmetric beam splitter, for $\beta=\sqrt{6}$, $\alpha=\sqrt{6}e^{-i\theta}$, and different values of $\gamma^{(3)}$, when the relative phase is set to $\theta=\pi/2$.
}
\end{figure*}

We express \begin{math} \ket{\psi} \end{math}, the wave function at the output of the beam splitter, in terms of combinations of Fock states \begin{math} \ket{n}_2\ket{m}_3 \end{math}, where 2,3 denote the output ports of the beam splitter. A Kerr state includes a nonlinear phase of the form \begin{math} exp[\gamma^{(3)}(\hat{n}^2-\hat{n})] \end{math} for every term in the coherent state (note that the linear part, \begin{math} \hat{n} \end{math} only adds a constant phase to the coherent state). \begin{math} \gamma^{(3)} \end{math} encompasses the (real) third-order optical susceptibility and the propagation length in the Kerr cell. 

Using the known relations between the input and output modes of a quantum beam splitter, the wavefunction at the output ports can be written in the following form:

\begin{eqnarray}
\ket{\Psi}=e^{-\frac{(|\alpha|^2+|\beta|^2)}{2}}\sum_{N=0}^{\infty} {\sum_{M=0}^{\infty}{\frac{\alpha^N\beta^M}{M!N!}}}e^{i\gamma^{(3)}(N^2+M^2)}
\nonumber\\\nonumber\\
\times
(t\hat{a}_3^\dagger{}+ir\hat{a}_2^\dagger{})^N(t\hat{a}_2^\dagger{}+ir\hat{a}_3^\dagger{})^M\ket{0}_2\ket{0}_3 \;\;,\;\;\;\;\;\;
\label{eq:Eq_1}
\end{eqnarray}
where $M$ and $N$ denote Fock states at the input ports, and $t$ and $r$ are the transmission and reflection coefficients, obeying $t^2+r^2=1$. From Eq. \ref{eq:Eq_1} we obtain an expression for the density matrix $\hat{\rho}_{23}=\ket{\psi}\bra{\psi}$ and then calculate the reduced density matrices for each output port by tracing over the states of the other port, namely $\hat{\rho}_{2}=Tr_3\{{\hat{\rho}_{23}}\}$ and $\hat{\rho}_{3}=Tr_2\{{\hat{\rho}_{23}}\}$. We then calculate the probabilities for measuring $n$ photons at port 2 and $m$ photons at port 3 by computing $\bra{n}\hat{\rho}_{2}\ket{n}$ and $\bra{m}\hat{\rho}_{3}\ket{m}$. The phase probability distributions can also be calculated from $\bra{\varphi}\hat{\rho}_{2}\ket{\varphi}$ and $\bra{\varphi}\hat{\rho}_{3}\ket{\varphi}$, where $\ket{\varphi}=\sum_ke^{ik\varphi}\ket{k}$, but here we focus on the photon number distribution.

The probability of measuring $x$ photons at port 2 is given by:
\begin{eqnarray}
p_2(x)=e^{-(|\alpha|^2+|\beta|^2)}\sum_{N=0}^{\infty} \sum_{M=0}^{\infty}\sum_{K=0}^{\infty}\sum_{m=0}^{M}\sum_{k=0}^{K}\beta^M(\beta^*)^K
\nonumber\\\nonumber\\
\times\alpha^N(\alpha^*)^Le^{i\gamma^{(3)}(M^2+N^2-K^2-L^2)}\;\;\;\;\;\;\;\;\;\;\;\;\;
\nonumber\\\nonumber\\\times(ir)^{M-m+n}(-ir)^{K-k+l}\cdot t^{N-n+m+k+L-l}\;\;\;\;\;\;\;\;\;
\nonumber\\\nonumber\\
\times\frac{j!\;\;x!}{m!(M-m)!n!(N-n)!k!(K-k)!l!(L-l)!}\;\;,\;\;\;\;\;\;
\label{eq:Eq_2}
\end{eqnarray}

where $j=M+N-m-n$, $x=m+n$, $l=x-k$ and $L=M+N-K$ (these relations are obtained from the delta functions involved in tracing and calculating the probabilities; $M$,$N$,$K$ and $L$ denote Fock states at the input ports, and $m$,$n$,$k$ and $l$ denote Fock states at the output ports). Similarly, the probability of measuring $y$ photons at port 3 is given by:
\begin{eqnarray}
p_3(y)=e^{-(|\alpha|^2+|\beta|^2)}\sum_{N=0}^{\infty} \sum_{M=0}^{\infty}\sum_{K=0}^{\infty}\sum_{m=0}^{M}\sum_{k=0}^{K}\beta^M(\beta^*)^K
\nonumber\\\nonumber\\
\times\alpha^N(\alpha^*)^Le^{i\gamma^{(3)}(M^2+N^2-K^2-L^2)} \;\;\;\;\;\;\;\;\;\;\;\;\;
\nonumber\\\nonumber\\\times(ir)^{M-m+n}(-ir)^{K-k+l}\cdot t^{N-n+m+k+L-l}\;\;\;\;\;\;\;\;\;
\nonumber\\\nonumber\\
\times\frac{j!\;\;y!}{m!(M-m)!n!(N-n)!k!(K-k)!l!(L-l)!}\;\;,\;\;\;\;\;\;
\label{eq:Eq_3}
\end{eqnarray}
\\where $j=m+n$, $n=M+N-m-y$, $l=j-k$ and $L=M+N-K$. Equations \ref{eq:Eq_2} and \ref{eq:Eq_3} allow us to calculate the photon number distributions at the two output ports when there are two Kerr states with the same absolute amplitudes at the two input ports, with a relative phase $\theta$. Without loss of generality, $\beta$ at port 0 can be chosen real, and the state at port 1 is then $\beta e^{-i\theta}$.

Before showing the photon number distributions, we demonstrate that our expressions conserve the total average number of photons (namely the total energy). Figure \ref{fig:n_average} presents the results of a numerical evaluation of Eq. \ref{eq:Eq_2} and Eq. \ref{eq:Eq_3}, for a 50:50 beam splitter, $\beta=2$ and $\gamma^{(3)}=0.1$. The total nonlinear phase shift in this case is comparable to experimentally attainable values when rubidium vapor (D2 line) is used as a nonlinear medium: for a 1 Watt laser beam with a moderate cross section area of $I=0.25\;cm^2$, a nonlinear index of refraction $n_2=1.5\cdot10^{-7}\;cm^2/W$ [\onlinecite{PhysRevA.69.023804}], and a propagation distance of $12.5\;cm$, the nonlinearity-induced path length difference is $\Delta n\cdot l =n_2I\cdot l=0.075\;\mu m$, resulting in a phase shift of $2\pi\cdot\frac{0.075}{0.78}\approx0.6\;rad$ at a wavelength of $780\;nm$; the corresponding value in the calculation is $\gamma^{(3)}\langle n^2\rangle=0.1\cdot6=0.6$. Figure
\ref{fig:n_average}(a) shows the calculated average photon numbers at the two output ports as a function of the relative phase between the inputs. The total average photon number is conserved, but the visibility of the interference fringes is reduced compared to interfering coherent states, as expected. Figure \ref{fig:n_average}(b) shows that the visibility of the interference fringes gradually falls to zero as $\gamma^{(3)}$ is increased.
\begin{figure}[t]
\includegraphics[width=8.6cm]{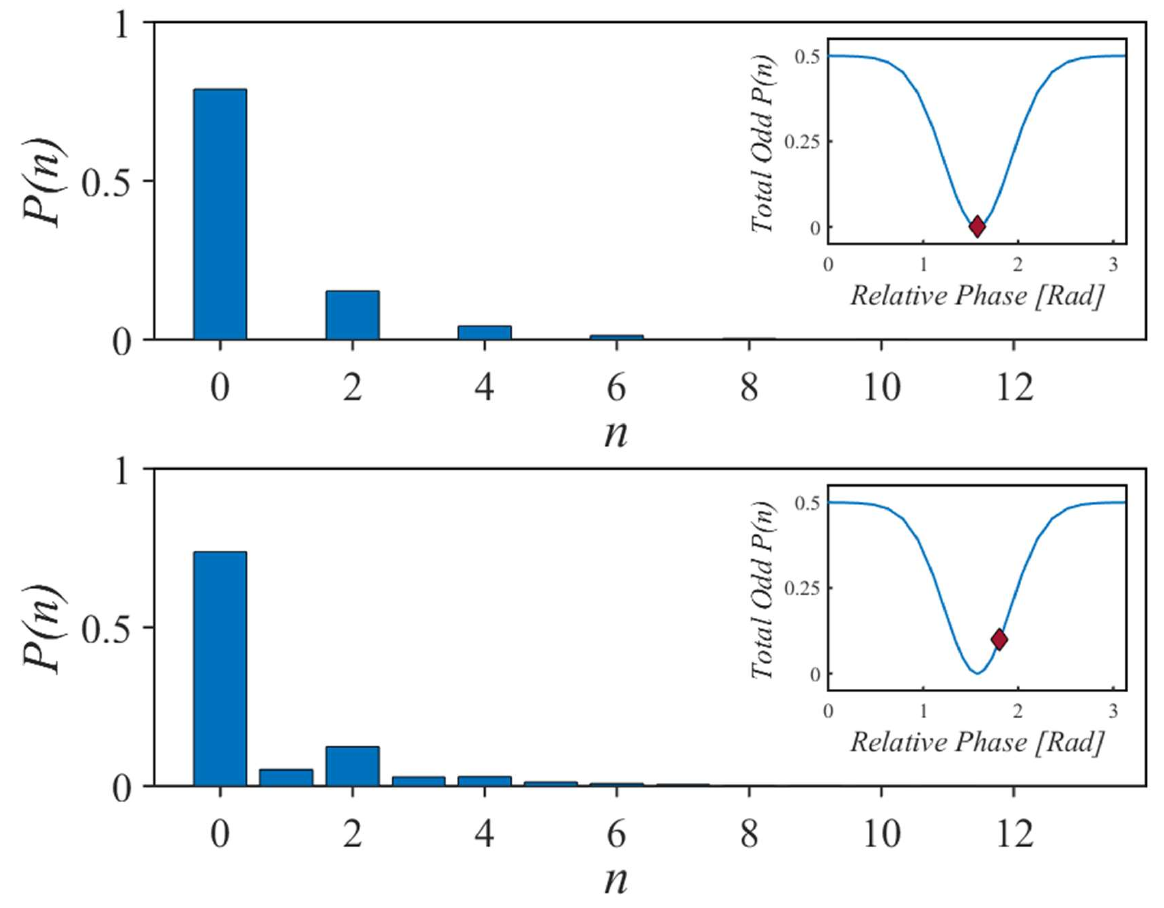}
\caption{\label{fig:around_half_pi} the calculated photon number distribution at output port 3 of a symmetric beam splitter, for $\beta=2$, $\alpha=2e^{-i\theta}$, and $\gamma^{(3)}=0.1$, when the relative phase is set to exactly $\theta=\pi/2$ (top), and slightly off the optimum point (bottom) – see insets.
}
\end{figure}

We now turn to the photon number distributions. Figure \ref{fig:with_increasing_gamma}(a) and Fig. \ref{fig:with_increasing_gamma}(b) show the calculated photon number distributions at the two output ports of the same symmetric beam splitter, for $\beta=\sqrt{6}$ and $\gamma^{(3)}=0.1$, when the relative phase is set to  $\theta=\pi/2$, so as to obtain an “almost dark” port 3 (see Fig. \ref{fig:n_average}). Port 3 then shows a clear odd-even effect, namely it is exclusively populated by even Fock states, while all odd photon number probabilities are zero. States such as this are more generally known as “photon number oscillations” [\onlinecite{schleich1987oscillations},\onlinecite{DODONOV1989211}], and can result from superpositions of coherent states (when they are known as Schrödinger cat states [\onlinecite{gerry2023introductory},\onlinecite{DODONOV1974597,PhysRevA.45.6570,doi:10.1080/09500349314551131}]), as
well as in correlated and squeezed light [\onlinecite{schleich1987oscillations},\onlinecite{DODONOV1989211},\onlinecite{PhysRevA.81.013814},\onlinecite{PhysRevLett.92.113602}]. They are a hallmark of squeezed vacuum [\onlinecite{gerry2023introductory},\onlinecite{breitenbach1997measurement},\onlinecite{PhysRevA.81.013814}], in which case they are a consequence of pair-wise generation of photons. To the best of our knowledge, such states have not been associated with Kerr squeezing before. While one type of Schrödinger cat state, the Yurke-Stoler state [\onlinecite{PhysRevLett.57.13}], \textit{can} be generated in a Kerr-like medium, it is not an even or odd cat state, and possesses a Poisson photon number distribution. In addition, Yurke-Stoler states evolve at specific times (or, equivalently, lengths of propagation in the nonlinear medium), while in our case the non-classical statistics emerge for arbitrary $\gamma^{(3)}$. Interference of identical Kerr states in a balanced beam splitter was proposed [\onlinecite{adam2011double}] as a conditional scheme for preparing optical Schrödinger cat states, but the effect of relative phase, and destructive interference in particular, was not explored. Moreover, the proposed scheme allowed the preparation of one of a set of Schrödinger-cat states only in the limit of a weak Kerr effect. In contrast, our scheme produces even-odd oscillations for arbitrarily large $\gamma^{(3)}$. This is demonstrated in panels (c)-(f) of Fig. \ref{fig:with_increasing_gamma}. In the low nonlinearity limit, the even-odd states that we discuss here resemble squeezed vacuum states, with a monotonously decreasing photon number probability, but the even-odd oscillations persist for higher nonlinearities, when the photon number distribution has a second peak at higher values of $n$ (in addition to the peak at $n$=0). The effect is robust in the sense that it washes out slowly as a function of the relative phase (see Fig. \ref{fig:around_half_pi}). $\gamma^{(3)}=0$ is the limit in which only the vacuum Fock state $\ket{0}$ is populated, as expected for a destructive interference of identical coherent states.

We now show that the even-odd effect reflects the inherent symmetry of the equations, which expresses itself in perfect pair-wise cancellation of amplitudes that contribute to the odd Fock states. For arbitrary $\beta, \alpha=\beta e^{-i\theta}, \theta=\pi/2$, and $r=t=1/\sqrt{2}$, Eq. \ref{eq:Eq_3} reduces to the following expression:
\\
\begin{eqnarray}
p_3(y)=e^{-(2|\beta|^2)}\sum_{N=0}^{\infty} \sum_{M=0}^{\infty}\sum_{K=0}^{\infty}\sum_{m=0}^{M}\sum_{k=0}^{K}|\beta|^{M+K+N+L}
\nonumber\\\nonumber\\
\times e^{i\gamma^{(3)}(M^2\;\;+N^2-K^2-L^2)} \;\;\;\;\;\;\;\;\;\;\;\;\;\;\;
\nonumber\\\nonumber\\\times(-i)^{N+K-k+l}(i)^{L+M-m+n}\cdot (\frac{1}{\sqrt{2}})^{M+K+N+L}\;\;
\nonumber\\\nonumber\\
\times\frac{j!\;\;y!}{m!(M-m)!n!(N-n)!k!(K-k)!l!(L-l)!}\;\;\;.\;\;\;\;
\label{eq:Eq_4}
\end{eqnarray}
\\
A simultaneous exchange $K\leftrightarrow L$ and $k\leftrightarrow l$ in Eq. \ref{eq:Eq_4} leaves the first two and last two terms unchanged, while $(-i)^{K-k+l}(i)^L$ 
transforms to $(-i)^{L-l+k}(i)^K$. It is straightforward to show that the ratio of these two expressions is $(-1)^{L-K+k-l}=(-1)^{2(L-l)-y}=(-1)^{-y}$. Thus, for any odd $y$ the two terms cancel out,
while for any even $y$ they add up. A simultaneous exchange $M\leftrightarrow N$ and $m\leftrightarrow n$ leads to the same result. This implies pair-wise cancellation of the terms contributing to $p_3(y)$ for any odd $y$ - an interference effect akin to the Hong-Ou-Mandel experiment.

In conclusion, we have demonstrated that destructive interference of identical Kerr-squeezed states leads to light with exclusively even Fock states. This is an interference effect that results from perfect pair-wise cancellation of amplitudes that contribute to the odd states. Its observability with an arbitrarily low nonlinearity is only limited by the quality of the interferometer. In the low nonlinearity limit the new state resembles a squeezed vacuum state. The even-odd oscillations persist when the nonlinearity is strong, in which case they are robust against deviations from the optimum relative phase of the input Kerr squeezed states.


\providecommand{\noopsort}[1]{}\providecommand{\singleletter}[1]{#1}%

\end{document}